\documentclass[10pt,fleqn,twocolumn]{article}
\usepackage[utf8]{inputenc}
\usepackage[english]{babel}
\pdfoutput=1
\usepackage{amsmath,amsfonts,amsthm,mathrsfs}
\usepackage[full]{textcomp}
\usepackage[protrusion=true,expansion=true]{microtype}	
\usepackage{graphicx,color}
\usepackage[a4paper,includeheadfoot,left=1.3cm,right=1.3cm,top=1cm,head=0.5cm,bottom=1.7cm,foot=0.7cm]{geometry}
\usepackage{booktabs}
\usepackage{enumitem}
\usepackage[version=4]{mhchem}
\mhchemoptions{minus-sidebearing-left=0.5em,minus-sidebearing-right=0.5em}
\usepackage{flushend, balance} 

\usepackage[T1]{fontenc}
\usepackage{lmodern}

\usepackage{siunitx}
\DeclareSIUnit\gmol{g\text{-}mol}
\DeclareSIUnit\kgmol{kg\text{-}mol}
\DeclareSIUnit\lbmol{lb\text{-}mol}
\DeclareSIUnit\molar{\mole\per\cubic\deci\metre}
\DeclareSIUnit\Molar{M}
\DeclareSIUnit\torr{torr}
\DeclareSIUnit\micron{\micro\metre}
\DeclareSIUnit\mrad{\milli\rad}
\DeclareSIUnit\gauss{G}
\DeclareSIUnit\rpm{rpm}
\DeclareSIUnit\inch{in}
\DeclareSIUnit\watt{W}
\DeclareSIUnit\ppm{ppm}
\DeclareSIUnit\sccm{sccm}

\usepackage[compress, numbers, super]{natbib}
\setcitestyle{round} 

\usepackage[font=small,labelfont={color=black,bf},textfont={color=black},format=plain,indention=0cm,justification=centerlast,skip=0.1cm]{caption,subfig}
\makeatletter
\renewcommand{\fnum@figure}{\textbf{\small\mbox{Fig.~\thefigure}}}
\renewcommand{\fnum@table}{\textbf{\small\mbox{Table~\thetable}}}
\makeatother

\newcommand{\rect}[1]{\mathrm{rect}{#1}}
\newcommand{\finofdoc}{\rule{0.09in}{0.09in}}
\newcommand{\mytitle}{\textbf{3-omega method for thermal properties of thin film multilayers}}
\newcommand{\linefindoc}{\color{black}
\vspace{0.5cm}
\centering\rule{0.7\linewidth}{1.2pt}\\%
\vspace{-0.39cm}
\rule{0.5\linewidth}{1.2pt}\\%
\vspace{-0.39cm}
\rule{0.3\linewidth}{1.2pt}%
\vspace{-0.5cm}}

\usepackage[colorlinks=true]{hyperref}
\hypersetup{
pdffitwindow=true,
pdftitle={\mytitle},
pdfauthor={L. N. Acquaroli},
pdfsubject={},
pdfcreator={L.N. Acquaroli},
pdfproducer={L.N. Acquaroli},
pdfkeywords={},
colorlinks={true},
linkcolor={NeonBlue},
citecolor={Cinnabar},
filecolor={Cinnabar},
urlcolor={NeonBlue}
}
\usepackage{url} 

\definecolor{NeonBlue}{rgb}{0.11,0.22,0.73} 
\definecolor{Cinnabar}{rgb}{0.8078,0.0863,0.1255}

\usepackage{titlesec}
\titlelabel{\thetitle. }
\titleformat*{\section}{\centering\large\bfseries}
\titleformat*{\subsection}{\centering\bfseries}

\usepackage{titling}	

\pretitle{\vspace{-40pt} \begin{center} \vspace{5pt} \fontsize{14}{14} \bfseries \color{black} \selectfont }
\title{\mytitle}
\posttitle{\par\end{center}\vspace{-.15cm}}
\preauthor{\vspace{-13pt} \begin{center}
\bigskip \color{black}}
\author{\textbf{Leandro N. Acquaroli}}	
\postauthor{\small \color{black}
\medskip

{Department of Engineering Physics, Ecole Polytechnique Montreal

P.O. Box 6079, Station Centre-Ville, Montreal (QC) H3C 3A7, Canada} 
\vspace{-0.4cm}
\date{\small\today}
\par\end{center}}

\usepackage{fancyhdr}	
\usepackage{lastpage}
\renewcommand{\headrulewidth}{0.0pt}
\renewcommand{\footrulewidth}{0.0pt}

\usepackage{abstract}
\usepackage{indentfirst}

\renewcommand{\thetable}{\Roman{table}}

\begin{document}

\setlength{\belowdisplayskip}{5pt}\setlength{\belowdisplayshortskip}{5pt}
\setlength{\abovedisplayskip}{5pt}\setlength{\abovedisplayshortskip}{5pt}

\columnsep 0.6cm

\renewcommand{\abstractname}{}
\twocolumn[
\maketitle
\vspace{-1.7cm}
\begin{onecolabstract}
Short review on the different models for the electro-thermal 3-omega method. We present the deduction of the fundamental relation between the $3\omega$ voltage with the temperature rise to determine the thermal conductivity. The usage of the anisotropy of the films allows a smooth transition between 1D and 2D models. A comparison between the multilayer methods and analytical solutions are presented.
\end{onecolabstract}
\vspace{1cm}
]

\fancypagestyle{plain}{%
\fancyhf{} 
\fancyfoot[R]{\footnotesize \thepage\ of \pageref{LastPage}} 
\renewcommand{\headrulewidth}{0pt}
\renewcommand{\footrulewidth}{0pt}}

\section{Introduction}

\begin{figure}[t]
    \begin{center}
        \includegraphics[scale=0.4]{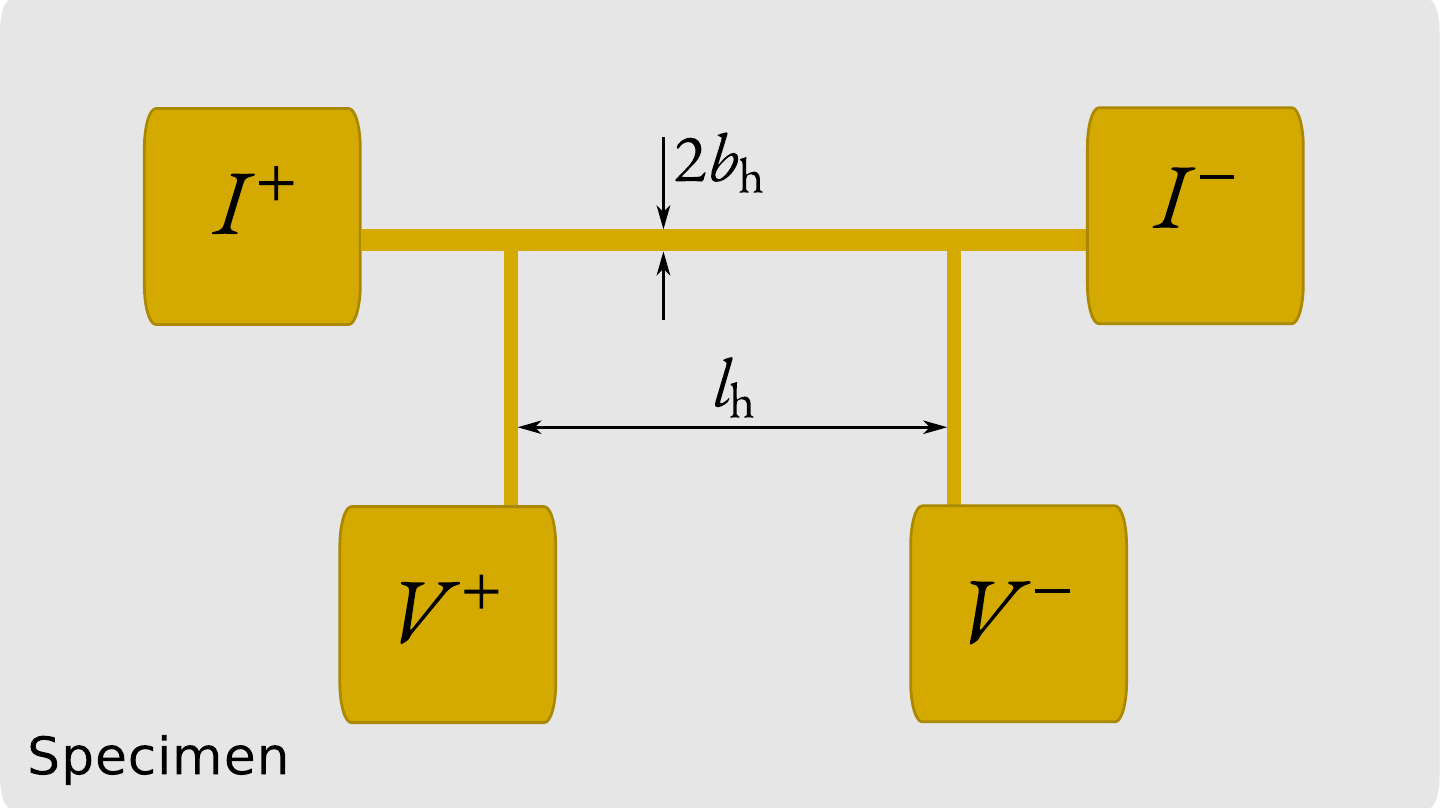}
        \vspace{0.3cm}
        \caption{Schematic of the metal line filament deposited on a specimen used for the $3\omega$ measurements. $l _{\text{h}}$ and $b_{\text{h}}$ denote the heater length and half-width, respectively. The contact pads are used to make electrical contact with the microfabricated heater, where the current and voltages inputs are depicted as well. An AC current source is thought for this.\hfill{ }}
        \label{fig_rtd}
        \vspace{-0.5cm}
    \end{center}
\end{figure}

In the 3-omega method, a metal line is deposited on the surface of the specimen of interest to serve as both the heating source and thermometer detector ---Fig.~\ref{fig_rtd}---. An alternating current of frequency $\omega$ is passed through the wire, $I_{1\omega}$, inducing a Joule heating event in the wire, $Q_{2\omega} = I_{1\omega}^2\,R_0$, where $R_0$ is the steady state electrical resistance of the wire. The heating event causes a second harmonic temperature rise in the sample, $T_{2\omega}$, which is a function of the thermophysical properties of the underlying sample. The electrical resistance of the wire varies linearly with respect to temperature ---$\mathit{\Delta} R/\mathit{\Delta} T$--- across moderate temperature ranges in which the metal does not undergo a phase change. This temperature varying resistance, $R_{2\omega}$, induced by $T_{2\omega}$ provides the third harmonic voltage oscillation, from which the experiment obtains its name, via Ohm's law: $V_{3\omega}=R_{2\omega}\times I_{1\omega}$. This quantity can be measured by a lock-in amplifier (which often serves as the function generator as well)~\cite{cahill1990, bauer2014}. For small temperature changes, the resistance of the filament varies with temperature as~\cite{koninck2008, haninnen2013}
\begin{equation}\label{eq_08}
R = R_0\,(1+\beta\,\mathit{\Delta} T)\,,
\end{equation}

\noindent
where $\beta$ is the temperature coefficient of resistance ---TCR---, and $R_0$ and $R$ are the resistances at temperatures $T_0$ and $T_0+\mathit{\Delta} T$ , respectively. The power dissipated by the heater/RTD ---from now on referred to simply as the heater--- due to Joule heating is defined as
\begin{equation}\label{eq_09}
P = I_{\text{h}}^2\, R_{\text{h}}\,,
\end{equation}

\noindent
where $I_{\text{h}}$ and $R_{\text{h}}$ are the heater current and resistance respectively. The alternating current passing through the heater is given as
\begin{equation}\label{eq_10}
I_{\text{h}}(t) = I_{\text{h},0}\,R_{\text{h},0}\,,
\end{equation}

\noindent
where $I_{\text{h},0}$ is the peak amplitude of the nominal heater current at a frequency $\omega$. Assuming that the change in resistance is negligible compared to the amplitude of the current, the instantaneous power can be written as
\begin{equation}\label{eq_11}
P(t) = \frac{1}{2}\,I_{\text{h},0}^2\,R_{\text{h},0}\,[1+\cos(2\omega t)]\,,
\end{equation}

\noindent
where $R_{\text{h},0}$ is the nominal heater resistance. Then, The power can be separated into two components: a constant component independent of time and an oscillating component:
\begin{align}
P_{\text{DC}} &= \frac{1}{2}\,I_{\text{h},0}^2\,R_{\text{h},0} = P_0\label{eq_12}\\
P_{\text{AC}}(t) &= \frac{1}{2}\,I_{\text{h},0}^2\,R_{\text{h},0}\,\cos(2\omega t)= P_0\,\cos(2\omega t)\label{eq_13}.
\end{align}

The average power dissipated by the heater is also called the root mean square (rms) power, which is half of the power dissipated by a DC current of the same amplitude, defined as
\begin{equation}\label{eq_14}
P_{\text{rms}} = I_{\text{h,rms}}^2\,R_{\text{h},0}=P_0\,,
\end{equation}

\noindent
where the rms heater current is given by
\begin{align}
I_{\text{h,rms}} &= \sqrt{\frac{1}{\tau}\int_0^\tau I_{\text{h}^2(t)\,\text{d}t}}\nonumber\\
&=I_{\text{h,0}}\sqrt{\frac{\omega}{2\pi}\int_0^{2\pi/\omega}\cos^2(\omega t)\text{d}t}\nonumber\\
&=\frac{I_{\text{h,0}}}{\sqrt{2}}\label{eq_15}\,.
\end{align}

Assuming that the heater circuit is stable, i.e., that all the transient perturbations decay over time, the steady-state harmonic temperature oscillations in the metal filament produce harmonic variations in the resistance given by
\begin{equation}\label{eq_16}
R_{\text{h}}(t) = R_{\text{h},0}\, [ 1 + \beta_{\text{h}}\, \mathit{\Delta} T_{\text{DC}} + \beta_{\text{h}}\, \mathit{\Delta} T_{\text{AC}}\, \cos(2\omega t + \phi) ]\,,
\end{equation}

\noindent
where $R_{\text{h},0}$ is the nominal ---room temperature--- resistance of the heater, $\mathit{\Delta} T_{\text{DC}}$ is the steady-state temperature increase due to the rms power dissipated by the filament, $\mathit{\Delta} T_{\text{AC}}$ is the magnitude of the steady-state temperature oscillations due to the sinusoidal component of the power and $\phi$ is the phase angle between the temperature oscillations and the excitation current. The resulting voltage across the sensor is obtained by multiplying the input current be the heater resistance yielding
\begin{align}
V_{\text{h}}(t) =  I_{\text{h},0}\, R_{\text{h},0}\, &\bigg[ (1 + \beta_{\text{h}}\, \mathit{\Delta} T_{\text{DC}})\cos(\omega t) \bigg.\nonumber\\
&+ \frac{1}{2} \beta_{\text{h}}\, |\mathit{\Delta} T_{\text{AC}}|\,\cos(\omega t +\phi)\nonumber\\
&\bigg.+ \frac{1}{2}  \beta_{\text{h}}\, |\mathit{\Delta} T_{\text{AC}}|\,\cos(3\omega t +\phi) \bigg]\,.\label{eq_17}
\end{align}

The $3\omega$ and $\omega$ components in the voltage with phase shift result from multiplying the $2\omega$ term in the resistance with the input current oscillating at the frequency $\omega$. We can express the $3\omega$ amplitude as amplitude as
\begin{equation}\label{eq_18}
V_{\text{h},3\omega} = \frac{1}{2}\,V_{\text{h},0}\,\beta_{\text{h}}\,|\mathit{\Delta} T_{\text{AC}}| = \frac{1}{2}\,V_{\text{h},0}\,\beta_{\text{h}}\,T_{2\omega}\,,
\end{equation}

\noindent
where $V_{\text{h},0} = I_{\text{h},0}R_{\text{h},0}$, and $T_{2\omega}$ is the complex AC temperature oscillations at $2\omega$. This is the fundamental result for the $3\omega$ method, which can be expressed in general terms of transfer functions~\cite{dames2005}, since by measuring the voltage at the frequency $3\omega$ one is able to deduce the in-phase ---$\Re(T_{2\omega})$--- and out-of-phase ---$\Im(T_{2\omega})$--- components of the temperature oscillations.

Both the magnitude and phase of the temperature oscillations vary with excitation frequency, due to the finite thermal-diffusion time ---$\tau_{\text{D}}$--- of the specimen. The thermal-diffusion time required for a thermal wave to propagate a distance $L$ is given by~\cite{birge1986}
\begin{equation}\label{eq_19}
\tau_{\text{D}} = \frac{L^2}{\alpha}\,,
\end{equation}

\noindent
where $\alpha$ is the thermal diffusivity, which is defined as the ratio of thermal conductivity ---$k$--- to volumetric heat capacity ---$c_p$---. In turn, the thermal-diffusion angular frequency is
\begin{equation}\label{eq_20}
\omega_{\text{D}} = \frac{2\pi}{\tau_{\text{D}}} = \frac{2\pi\alpha}{L^2}\,.
\end{equation}

In the limit of infinite thermal diffusivity ---i.e., infinite thermal conductivity---, heat propagates with infinite velocity such that the temperature is constant throughout the specimen. This leads to undamped temperature oscillations with zero phase lag. Conversely, zero thermal diffusivity (i.e., infinite volumetric heat capacity), results in no heat propagation, zero oscillation amplitude, and large phase lag. Note that the dissipation of the rms power is independent of the frequency response of the specimen since it is constant with time.

\section{Heat diffusion from an infinite planar heater with sinusoidal heating}

The sinusoidal current with frequency $\omega$ passing through the heater results in a steady DC temperature rise and an oscillating AC temperature component~\cite{haninnen2013}. The DC component creates a constant temperature gradient into the specimen while the AC part results in thermal waves diffusing into the specimen at frequency $2\omega$. Solving the heat equation in planar coordinates permits to determine the lag $\phi$ between the heater temperature and $T_{2\omega}$.

\begin{figure}[t]
    \begin{center}
        \includegraphics[scale=1.2]{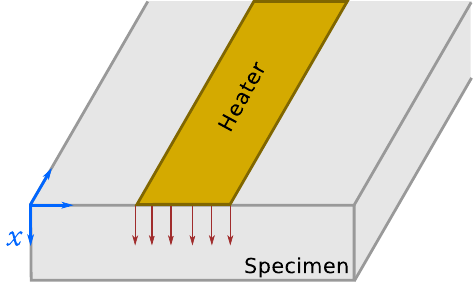}
        \vspace{0.3cm}
        \caption{One dimensional heat flow from a thin infinitely long plane heater on a substrate. Heat enters the specimen uniformly over the width of the heater and edge-effects are not considered. The zero
level of x-coordinate is located at the heater/specimen interface.\hfill{ }}
        \label{fig_planar_heater}
        \vspace{-0.5cm}
    \end{center}
\end{figure}

Consider Fig.~\ref{fig_planar_heater}. The heat diffusion equation to be solved for the system reads
\begin{equation}\label{eq_21}
T_{xx}-\frac{1}{\alpha}T_{t}=0\,.
\end{equation}

\noindent
In the case of sinusoidal heating the temperature can be separated into time and spatially dependent parts
\begin{equation}\label{eq_22}
T(x,t) = \mathit{\Theta}(x)\,\exp(i2\omega t)\,.
\end{equation}

\noindent
Replacing \eqref{eq_22} into \eqref{eq_21}, performing differentiation and dividing by $\exp(i2\omega t)$ yields:
\begin{equation}\label{eq_23}
\frac{\text{d}^2\mathit{\Theta}}{\text{d}x^2}-\frac{i 2\omega}{\alpha}\,\mathit{\Theta}=0\,.
\end{equation}

\noindent
The solution of \eqref{eq_23} can be expressed as
\begin{equation}\label{eq_24}
\mathit{\Theta}(x)=\mathit{\Theta}_0\exp(-qx)\,,\quad \text{for } x>0\,,
\end{equation}

\noindent
where $q$ is the thermal wavenumber defined as
\begin{equation}\label{eq_25}
q = \sqrt{\frac{i2\omega}{\alpha}}=(1+i)\sqrt{\frac{\omega}{\alpha}}=\sqrt{\frac{2\omega}{\alpha}}\exp(i\pi/4)\,.
\end{equation}

\noindent
Plugging \eqref{eq_23} and \eqref{eq_24} into \eqref{eq_22} gives
\begin{equation}\label{eq_26}
T(x,t) = \mathit{\Theta}_0\,\exp(i2\omega t-qx)\,.
\end{equation}

\noindent
Consider one of the boundary condition to be
\begin{align}
\mathit{\Phi}|_{x\to 0^+} &= -k\frac{\text{d}T}{\text{d}x}\bigg\vert_{x\to 0^+}\nonumber\\
&=kq\mathit{\Theta}_0\exp(i2\omega t)\nonumber\\
&=k\mathit{\Theta}_0\exp(i2\omega t-i\pi/4)\label{eq_27}\,.
\end{align}

\noindent
The flux equals the oscillating heat component produced in the heater per unit area. The oscillating power of the heater can be written in complex notation, according to \eqref{eq_13}:
\begin{equation}\label{eq_28}
P_0\,\cos(2\omega t) = P_0\,\Re[\exp(i2\omega t)]\,.
\end{equation}

\noindent
Dividing the complex power by the heater area $A$, setting it equal to the heat flux from \eqref{eq_27} and dividing both sides by $\exp(i2\omega t)$ gives
\begin{equation}\label{eq_29}
\mathit{\Theta}_0 = \frac{P_0}{Ak}\sqrt{\frac{\alpha}{2\omega}}\exp(-i\pi/4)\,.
\end{equation}

Finally, replacing this last result into \eqref{eq_26} produces
\begin{equation}\label{eq_30}
T_{2\omega}(x,t) = \frac{P_0}{Ak}\sqrt{\frac{\alpha}{2\omega}}\exp(i2\omega t-qx-i\pi/4)
\end{equation}

\noindent
establishing that the temperature rise lags the heater current by $\phi=\pi/4$, with a $\omega^{-1/2}$ dependence.

\section{Heat diffusion from a line heater inside a cylindrical specimen}

\begin{figure}[t]
    \begin{center}
        \includegraphics[scale=0.7]{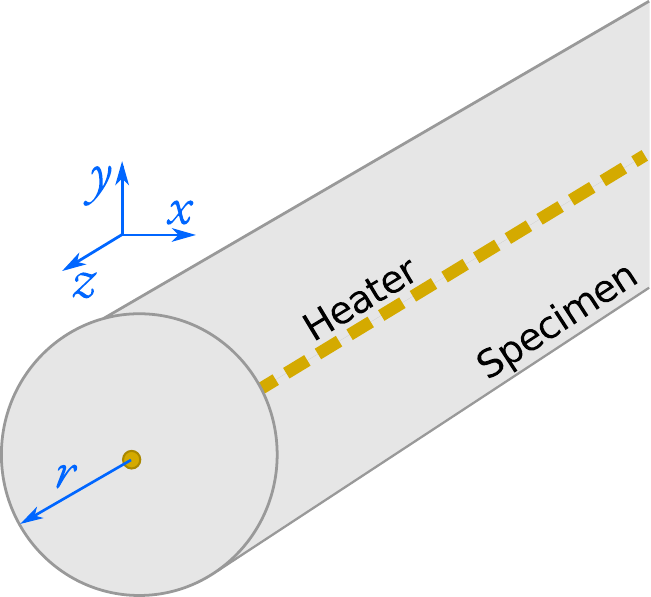}
        \vspace{0.3cm}
        \caption{The geometry of an infinite circular cylinder with one dimensional line heater running through the cylinder axis.\hfill{ }}
        \label{fig_cylinder_heater}
        \vspace{-0.5cm}
    \end{center}
\end{figure}

To understand how to use the $3\omega$ method to determine the thermal conductivity of a specimen, it is convenient to start by solving the heat equation of an infinitely long cylinder, when heated from an inside linear heater at a distance $r=\sqrt{x^2+y^2}$ ---Fig.~\ref{fig_cylinder_heater}---. The volume of the specimen is considered to be infinite, with no axial or circumferential temperature gradients.

The heat equation to be solved for this system reads as follow~\cite{carslaw1959}:
\begin{equation}\label{eq_31}
T_{rr} + \frac{1}{r} T_r - \frac{1}{\alpha} T_t= 0\,.
\end{equation}

\noindent
Given that the heating power is sinusoidal, the solution for this equation can be written as
\begin{equation}\label{eq_32}
T(r,t) = \mathit{\Theta}(r)\exp(i2\omega t)\,.
\end{equation}

\begin{figure}[t]
    \begin{center}
        \includegraphics[scale=0.47]{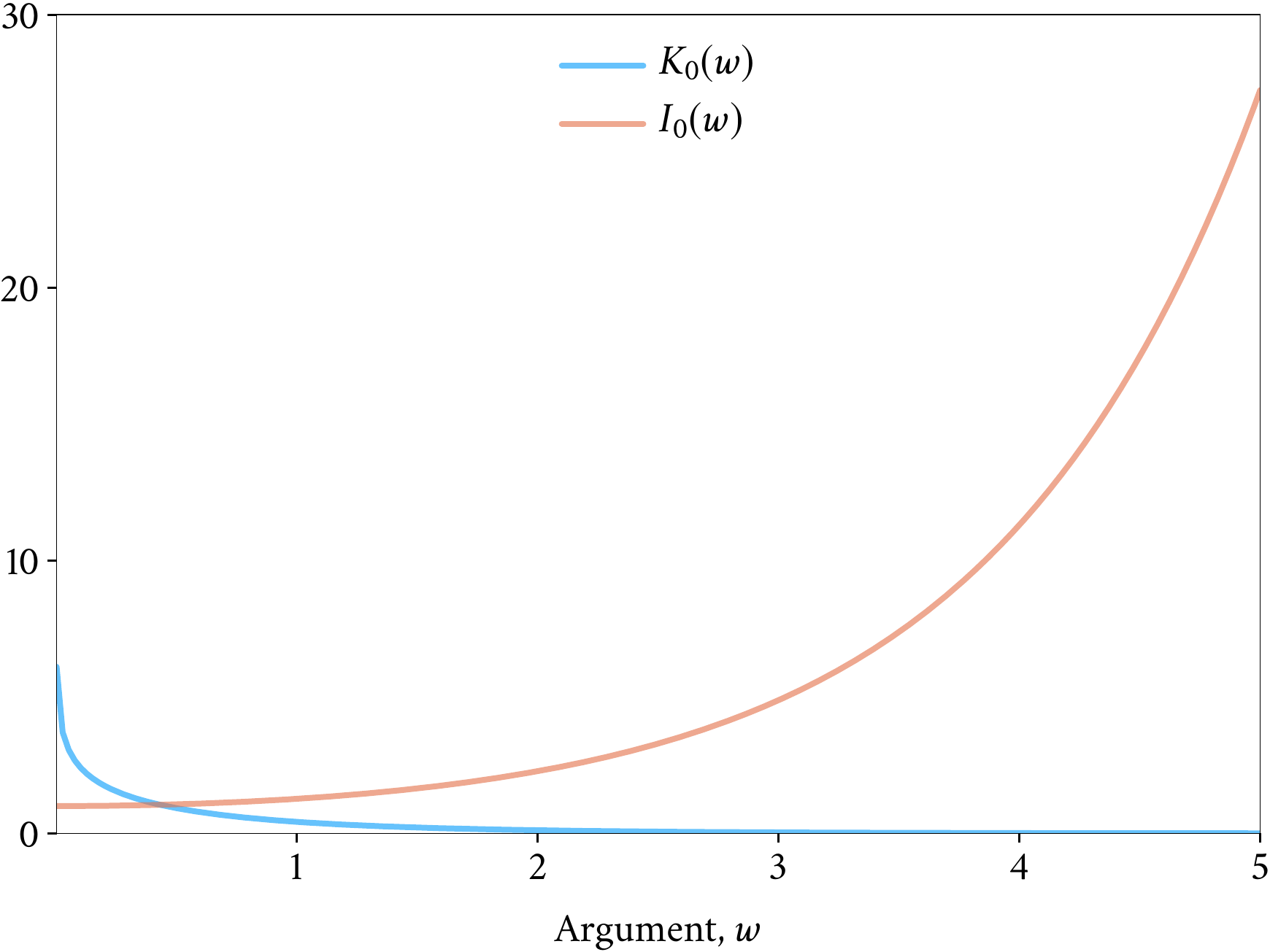}
        \vspace{0.3cm}
        \caption{Behavior of the Bessel functions.}
        \label{fig_bessel_functions}
    \end{center}
\end{figure}

\noindent
Introducing \eqref{eq_32} into \eqref{eq_31}, performing differentiation, and dividing by $\exp(i2\omega)$, gives
\begin{equation}\label{eq_33}
\frac{\text{d}^2\mathit{\Theta}}{\text{d}{r}^2} + \frac{1}{r} \frac{\text{d}\mathit{\Theta}}{\text{d}r} - q^2 \mathit{\Theta}= 0\,,
\end{equation}

\noindent
where $q^2=i2\omega/\alpha$, is the thermal wavenumber. Performing change of variables $w=qr$, leads to the zero-th order modified Bessel differential equation,
\begin{equation}\label{eq_34}
w^2\frac{\text{d}^2\mathit{\Theta}}{\text{d}{w}^2} + w \frac{\text{d}\mathit{\Theta}}{\text{d}w} - w^2 \mathit{\Theta} = 0\,,
\end{equation}

\noindent
whose solution can be written as
\begin{equation}\label{eq_35}
\mathit{\Theta}(w) = c_1 I_0(w) + c_2 K_0(w)\,,
\end{equation}

\noindent
where $I_0(w)$ is the zero-th order modified Bessel function of the first kind, and $K_0(w)$ is the zero-th order modified Bessel function of the second kind. The temperature profile of the system it is supposed to decay as $r\to\infty$, then it is required to set $c_1=0$ ---Fig.~\ref{fig_bessel_functions}---. Then, the second boundary condition can be written in terms of the heat flux at $r = r_0$:
\begin{align}
\mathit{\Phi}|_{r=r_0} &= -k\frac{\text{d}\mathit{\Theta}}{\text{d}r}\bigg\vert_{r=r_0}\nonumber\\
&=-c_2 k \frac{\text{d}\mathit{K_0(qr)}}{\text{d}r}\bigg\vert_{r=r_0}\nonumber\\
&=c_2 k q K_1(qr_0)\label{eq_36}\,.
\end{align}

On the other hand, the heat flux from the heater through a cylindrical surface at $r = r_0$ is related to the power and the cylinder area $A_{\text{cyl}}$ by
\begin{equation}\label{eq_37}
\mathit{\Phi}|_{r=r_0} = \frac{P}{A_{\text{cyl}}} = \frac{P_0}{2\pi r_0 l_{\text{h}}}\exp(i2\omega t)\,.
\end{equation}

\noindent
Equating \eqref{eq_36} and \eqref{eq_37}, yields
\begin{equation}\label{eq_38}
c_2 = \frac{P_0}{2\pi k l_{\text{h}}}\frac{1}{qr_0 K_1(qr_0)}=\frac{P_0}{2\pi k l_{\text{h}}}\,,
\end{equation}

\noindent
since $wK_1(w)\to 1$ when $w\to 0$. Then,
\begin{equation}\label{eq_39}
\mathit{\Theta} = \frac{P_0K_0(qr)}{2\pi k l_{\text{h}}}\,.
\end{equation}

\noindent
Plugging last equation into \eqref{eq_32}, gives
\begin{equation}\label{eq_40}
T_{2\omega}(r,t) = \frac{P_0K_0(qr)}{2\pi k l_{\text{h}}}\exp(i2\omega t)\,.
\end{equation}

\section{One-dimensional line heater at the surface of the specimen}

By halving the cylinder of Fig.~\ref{fig_cylinder_heater} the heater remains at the surface of the specimen. The result for infinite volume can used to express the temperature oscillations for infinite half-volume. Halving the volume means that twice as much of heat flux flows into the remaining volume, assuming no radiation, convection or conduction on the surface. Thus, \eqref{eq_40} becomes
\begin{equation}\label{eq_41}
T_{2\omega}(r,t) = \frac{p_0K_0(qr)}{\pi k}\exp(i2\omega t)\,.
\end{equation}

\begin{figure}[t]
    \begin{center}
        \hspace{-1.15cm}\includegraphics[scale=0.46]{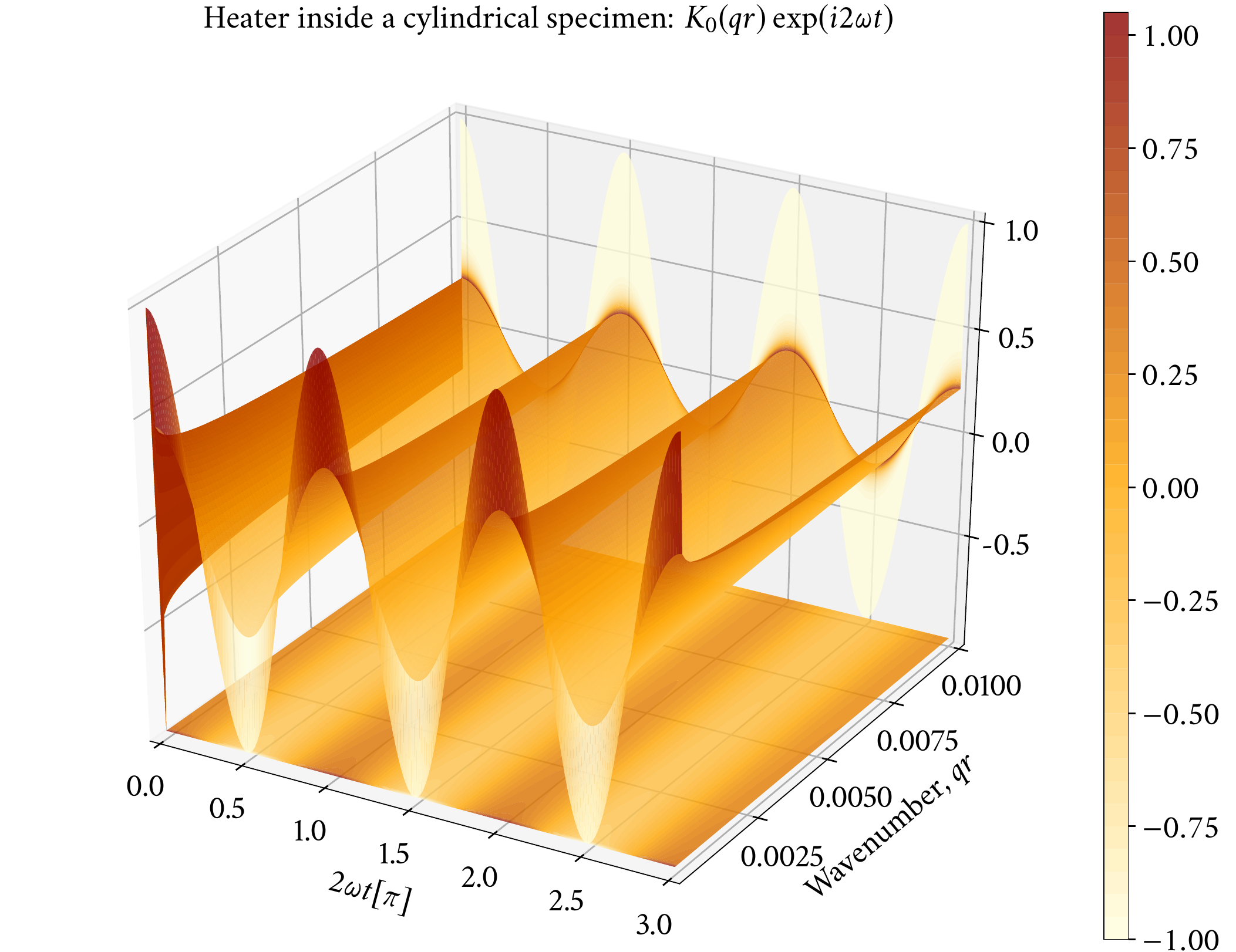}
        \vspace{0.2cm}
        \caption{Visualization of the normalized real part of equation~\eqref{eq_41}. The oscillations decay rapidly away from the heater and are even in the time-domain.\hfill{ }}
        \label{fig_t2w_cylinder}
        \vspace{-0.5cm}
    \end{center}
\end{figure}

\noindent
In Fig.~\ref{fig_t2w_cylinder} it is depicted the normalized real part of $T_{2\omega}\pi k/p_0$, where $p_0$ is the power per unit length of the heater.

The physical specimen may not be a semi-infinite solid, but rather has a finite thickness. Since the thermal waves decay rapidly away from the heat source ---following a modified Bessel function of the second kind--- it must
ensured that the penetration depth of the thermal wave is much less than the thickness of the specimen ---$\lambda<<d$---, where the thermal penetration depth is defined as
\begin{equation}\label{eq_43}
\lambda=\frac{1}{|q|}=\sqrt{\frac{\alpha}{2\omega}}\,.
\end{equation}

Since the amplitude of an exponentially decaying function reduces to approximately 1\% of its initial magnitude after 5 “length constants”, it is expected that the magnitude of the thermal oscillations will decrease below 1\% of its initial amplitude after 5 thermal penetration depths ---as the Bessel function decays faster than an exponential
function---. Thus, the specimen thickness must exceed 5 times the thermal penetration depth to be considered semi-infinite ---$d>>5\lambda$---.

\subsection{Effects of Finite Width of the Heater}

\begin{figure}[t]
    \begin{center}
        \includegraphics[scale=0.6]{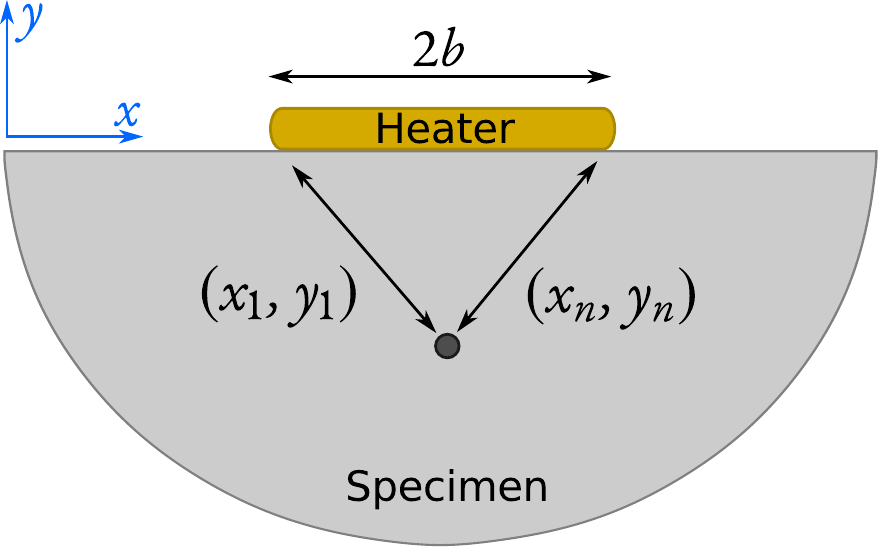}
        \vspace{0.3cm}
        \caption{Schematic of a finite-width ---$2b$--- line heater showing the temperature contributions from the extremities of the heater to a point in the specimen. \hfill{ }}
        \label{fig_linear_heater_superposition}
        \vspace{-0.5cm}
    \end{center}
\end{figure}

In order to generalize the 1D line solution to one with a finite-width, one needs to superimpose an infinite number of 1D line sources over the width of the heater~\cite{cahill1990}. This leads to a convolution integral, since the temperature at any arbitrary set of coordinates $(x,y)$ within the specimen depends on its relative position to all the line sources ---Fig.~\ref{fig_linear_heater_superposition}---. Mathematically this is done by taking a Fourier transform of \eqref{eq_41} with respect to x-coordinate ---as the temporal dependence is not of interest at this point---. Only the oscillations at the surface are important, so $y = 0$ and $r=x$. Now Fourier cosine transformation can be used because the temperature function is an even function. The Fourier cosine transform
pair is:
\begin{align}
\hat{f}(\eta)&=\int_0^\infty f(x) \cos(\eta x)\text{d}x\label{eq_43}\,,\\
f(x) &= \frac{2}{\pi} \int_0^\infty \hat{f}(\eta) \cos(\eta x)\text{d}\eta\label{eq_44}\,.
\end{align}

For the spatial temperature oscillations the cosine transform reads
\begin{align}
T_{2\omega}(\eta)&=\int_0^\infty T_{2\omega}(x) \cos(\eta x)\text{d}x\nonumber\\
&= \frac{p_0}{\pi k} \int_0^\infty K_0(q\eta) \cos(\eta x)\text{d}\eta\nonumber\\
&= \frac{p_0}{\pi k}\frac{1}{\sqrt{\eta^2+q^2}}\label{eq_45}\,.
\end{align}

The finite width can be added into \eqref{eq_45} by multiplying it with the Fourier transform of the heat source as a function of the x-coordinate. Heat enters the specimen evenly over the width of the heater ranging from $-b$ to $b$. This behavior can be expressed as a rectangular function with values 1 for $x < |b|$ and 0 elsewhere. According to the convolution theorem, the multiplication of the Fourier transforms of \eqref{eq_45} and the rectangular function equals the Fourier transform of the convolution of the functions in the x-space:
\begin{align}
T_{2\omega}(\eta)&= \frac{p_0}{\pi k}\frac{1}{\sqrt{\eta^2+q^2}} \int_0^\infty \rect(x) \cos(\eta x)\text{d}x
\nonumber\\
&=\frac{p_0}{\pi k}\frac{\sin(\eta b)}{\sqrt{\eta^2+q^2}}\label{eq_46}\,.
\end{align}

\noindent
Taking the inverse transform gives
\begin{equation}\label{eq_47}
T_{2\omega}(x)= \frac{p_0}{\pi k}\int_0^\infty \frac{\sin(\eta b)\cos(\eta x)}{(\eta b)\sqrt{\eta^2+q^2}}\text{d}\eta\,.
\end{equation}

Equation \eqref{eq_47} gives the form of temperature oscillations on the surface of the specimen a distance $x$ away from the center of the heat source that has a finite width $2b$. Since the thermometer and the heater are the same element the measured temperature is some average temperature over the width of the line. Expression~\eqref{eq_47} can be averaged by integrating it with respect to $x$ from 0 to $b$ and dividing by $b$ to give the temperature measured by the thermometer as proposed by Cahill~\cite{cahill1990}:
\begin{equation}\label{eq_48}
T_{2\omega}= \frac{1}{b}\int_0^b T_{2\omega}(x) \text{d}x=\frac{p_0}{\pi k}\int_0^\infty \frac{\sin^2(\eta b)}{(\eta b)^2\sqrt{\eta^2+q^2}}\text{d}\eta\,.
\end{equation}

\noindent
The asymptotic behavior of the last equation when $|q|b<<1$ ---normally termed slope-method for its dependence on $\ln(\omega)$--- can be written as~\cite{borca2001}
\begin{equation}\label{eq_48a}
T_{2\omega}= -\frac{p_0}{\pi k} \biggl[\ln\biggl(\frac{\alpha}{b^2}\biggr)-\frac{1}{2}\ln(\omega)+\zeta\biggr]-i\frac{p_0}{4k}\,.
\end{equation}

\noindent
where $\zeta$ is a constant. By defining the reduced frequency, $\mathit{\Omega}=b^2\omega/\alpha$, equation~\eqref{eq_48} results~\cite{duquense2010}:
\begin{equation}\label{eq_49a}
T_{2\omega}= \frac{p_0}{\pi k} \int_0^\infty \frac{\sin^2(\xi)}{\xi^2\sqrt{\xi^2+i\Omega}}\text{d}\xi\,.
\end{equation}

\noindent
With proper substitutions and calculations, the last expression is related to the Meijer G-function by~\cite{duquense2010}:
\begin{equation}\label{eq_49}
T_{2\omega}= \frac{-i\,p_0}{4\pi k \mathit{\Omega}} G_{24}^{22}\left(i\mathit{\Omega}\left|^{1,3/2}_{1,1,1/2,0}\right.\right)\,,
\end{equation}

\noindent
Asymptotic analysis gives two distinct regimes:
\begin{align}
T_{2\omega}|_{\mathit{\Omega\to 0}}&= \frac{p_0}{\pi k}\biggl(\frac{1}{2}\ln{\mathit{\Omega}}+\frac{3}{2}-\gamma-i\frac{\pi}{4}\biggr)\label{eq_50}\,,\\
T_{2\omega}|_{\mathit{\Omega\to\infty}}&=\frac{p_0}{2k\sqrt{2\mathit{\Omega}}}(1-i)\label{eq_51}\,,
\end{align}

\noindent
where $\gamma=0.5772$ is the Euler constant.

\subsection{Thin film on a substrate}

The effect of the thin film on the substrate can be added as a thermal resistance independent of the driving frequency. In the $3\omega$ method, the temperature drop across the film is inferred from the difference between the experimental temperature rise of the heater and the calculated temperature rise of the bare substrate sample. The substrate thermal conductivity is determined from the slope method applied to the experimental temperature rise of the heater and usually a one-dimensional heat conduction model is assumed across the film. With these approximations, the experimental temperature rise of the heater on the film plus substrate system can be written as~\cite{borca2001}
\begin{equation}\label{eq_52}
T_{\text{s+f},2\omega}=T_{\text{s},2\omega}+\frac{p_0\,d_{\text{f}}}{2 b k_{\text{f}}}\,,
\end{equation}

\noindent
where subscript ``f'' denotes film properties ---e.g. $d_{\text{f}}$ is the thickness of the film--- and $T_{\text{s},2\omega}$ is calculated based on Eq.~\eqref{eq_48a}. Considering the expression from Eq.~\eqref{eq_52}, the film thermal conductivity as determined from the slope-based $3\omega$ method may be written as
\begin{equation}\label{eq_53}
k_{\text{f}}=\frac{p_0\,d_{\text{f}}}{2 b}(T_{\text{f},2\omega}-T_{\text{s},2\omega})^{-1}_{\text{avg}}\,,
\end{equation}

\noindent
where $T_{\text{f},2\omega}$ is calculated from Eq.~\eqref{eq_18} with the measured voltages and TCR, and the average over the frequency range is performed. In order to determine the thermal conductivity of the film, the calculated temperature drop across the film is averaged over the experimental frequency range. If the system contains additional films, the thickness and thermal properties of these films must be known in order to
subtract their contribution from the total experimental temperature rise. Another way to calculate the thermal conductivity of the film may be carried out with the so-called differential $3\omega$ technique, where the thermal conductivity of the film is calculated using the average temperature rise difference experimentally measured at the same power input by similar heaters deposited on the specimen and a reference sample without the film of interest~\cite{borca2001}:
\begin{equation}\label{eq_54}
k_{\text{f}}=\frac{d_{\text{f}}}{2}\biggl[\biggl(\frac{b\,T_{2\omega}}{p_0}\biggr)_{\text{r+f}}-\biggl(\frac{b\,T_{2\omega}}{p_0}\biggr)_{\text{r}}\biggr]^{-1}_{\text{avg}}\,,
\end{equation}

\noindent
where subscript ``r+f'' corresponds to the structure that includes the sample film and the subscript ``r'' refers to the reference structure only. Although the differential technique requires two sets of measurements, the film thermal conductivity determined from Eq.~\eqref{eq_54} is insensitive to the substrate thermal conductivity, and the contributions from any additional films is nearly eliminated by the measurement on the reference sample. For further details and considerations on the applicability of the slope and differential methods, Borca-Tarsiuc 2001~\cite{borca2001} and Dames 2013~\cite{dames2013} are recommended readings.

\section{Heat conduction across a 2D multilayer-film on a substrate system}

The heat-conduction model developed by Borca-Tarsiuc was employed for the data analysis throughout the present work. Shortly, the model presents a solution based on a two-dimensional heat-conduction model across a multilayer system and a uniform heat flux boundary condition between the heater and the top film. Disregarding the contribution of the thermal mass of the heater, the complex temperature oscillations of a heater is given by~\cite{borca2001}:
\begin{equation}\label{eq_55}
T_{2\omega} = - \frac{p_0}{\pi\, k_{\text{y},1}}\, \int_{0}^{\infty} \frac{1}{\mathcal{A}_{1}(\eta)\, \mathcal{B}_{1}(\eta)}\frac{\sin^{2}(b\eta)}{(b\eta)^2 } \text{d}\eta
\end{equation}

\noindent
where
\begin{subequations}\label{eq_56}
\begin{align}
\mathcal{A}_{j-1} &= \frac{ \mathcal{A}_{j}\, \dfrac{ k_{\text{y},j}\, \mathcal{B}_{j} }{ k_{\text{y},j-1}\, \mathcal{B}_{j-1} } - \tanh(\mathcal{B}_{j-1}\, d_{j-1})}{ 1 - \mathcal{A}_{j}\, \dfrac{ k_{\text{y},j}\, \mathcal{B}_{j} }{ k_{\text{y},j-1}\, \mathcal{B}_{j-1} }\, \tanh(\mathcal{B}_{j-1}\, d_{j-1})}\,,\label{eq_56_1}\\
j&=2,\ldots,n\,,\nonumber\\
\mathcal{B}_{j} &= \sqrt{k_{\text{xy},j}\, \eta^2 + \frac{i2\omega}{\alpha_{\text{y},j}} }\,, \label{eq_56_2} \\
k_{\text{xy}} &= \frac{k_{\text{x}}}{k_{\text{y}}}\,, \label{eq_56_3}\,\\
\mathcal{A}_{n} &= -\tanh(\mathcal{B}_{n}\,d_n)^s\,. \label{eq_56_4}
\end{align}
\end{subequations}

In the previous expressions, $\eta$ is the integration variable, $i = \sqrt{-1}$, $n$ is the total number of layers including the substrate, $j$ corresponds to the $j$th layer starting from the top, subscript y denotes the cross-plane perpendicular to the film$/$substrate interface, while x defines the in-plane coordinate, $b$ and $l$ are the half-width and length (where the voltage drop is read) of the heater, respectively, $p$ is the peak electrical power per unit length, $k$ is the thermal conductivity, $\alpha=k/(\rho\,c_{\text{p}})$ the thermal diffusivity (where $\rho$ is the density and $c_{\text{p}}$ is the specific heat capacity of the layer), $\omega$ indicates the angular frequency and $d$ the thickness of the layer. Equation~\eqref{eq_56_4} establishes the substrate boundary condition, where $s$ takes the values 0 for semi-infinite, 1 for an adiabatic condition and -1 for an isothermal boundary when the substrate is finite.

\subsection{Matrix formalism}

The matrix formalism is similar to that reported elsewhere~\cite{bauer2014, borca2001, kim1999, olson2005}. Consider a multilayer system in vacuum with the layers numbered from bottom to the top, 1 to $n$, where the heater is in contact with the top layer. The $3\omega$ signal is obtained from the ac temperature of the top surface of the top layer. An ac current at $\omega$ flows through the heater strip normal to the plane of page, generating heat at $2\omega$. The ac surface temperature for a multilayer system averaged over the strip, is thus~\cite{kim1999}:
\begin{multline}\label{eq_57}
T_{2\omega} = \frac{p_0}{2\pi}\times\\
\int_{0}^{\infty}\frac{\mathcal{B}^+(\eta)+\mathcal{B}^-(\eta)}{\mathcal{A}^+(\eta)\,\mathcal{B}^-(\eta)-\mathcal{A}^-(\eta)\,\mathcal{B}^+(\eta)}\,\frac{\sin^2(\eta b)}{\lambda_{\text{h}}\,(\eta b)^2}\, \text{d}\eta
\end{multline}

\begin{figure}[t]
    \begin{center}
        \resizebox{\columnwidth}{!}{\includegraphics{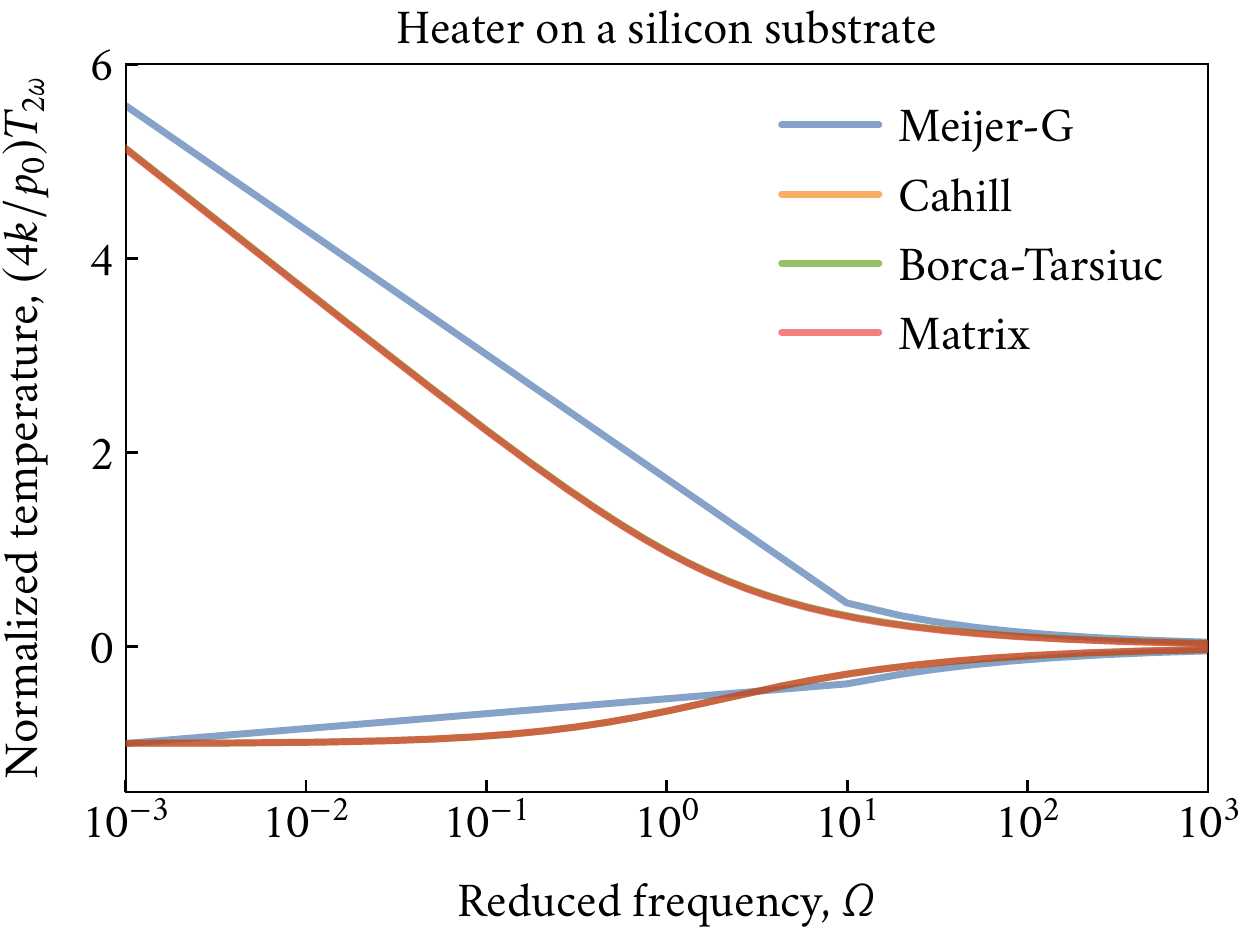}}
        \caption{Normalized $T_{2\omega}$ spectrum of isotropic bulk silicon with $k_y=\SI{160}{\watt\per\meter\per\kelvin}$, $l=\SI{1}{\milli\meter}$, $b=\SI{6.25}{\micro\meter}$, $p=\SI{0.02}{\watt}$, calculated with Eqs.~\eqref{eq_49a} ---Cahill---, \eqref{eq_49} ---Meijer-G---, \eqref{eq_55} ---Borca-Tarsiuc--- and \eqref{eq_57} ---Matrix formalism---.\hfill{ }}
        \label{fig_2}
    \end{center}
\end{figure}

\noindent
with
\begin{equation}\label{eq_58}
\lambda_j = k_{\text{y},j}\,\sqrt{k_{xy,j}\,\eta^2 + i\frac{2\omega}{\alpha_{y,j}}},\quad k_{xy,j} = \frac{k_{x,j}}{k_{y,j}}\,,
\end{equation}

\noindent
where the subscript h ---first layer, $j=n$--- refers to the heater metal strip, $\mathcal{A}^+(\eta)$, $\mathcal{A}^-(\eta)$, $\mathcal{B}^+(\eta)$, and $\mathcal{B}^-(\eta)$ are parameters determined by a matrix procedure analogous to that given in reference~\cite{kim1999}. The recursion relations show symbolically how $\mathcal{B}^+(\eta)$ and $\mathcal{B}^-(\eta)$ change if we add a layer, considering the thermal resistance of the interfaces~\cite{bauer2014}. For instance, since the semi-infinite boundary condition ---BC--- applies at the bottom of the substrate, for a heater on a substrate system we have:
\begin{equation}\label{eq_59a}
\begin{pmatrix} \mathcal{B}^+\\\mathcal{B}^-\end{pmatrix}=\underbrace{\frac{1}{2\lambda_{2}}\begin{pmatrix} \lambda_{2}+\lambda_1+\lambda_{2}\lambda_1 R_1 & \lambda_{2}-\lambda_1-\lambda_{2}\lambda_1 R_1\\\lambda_{2}-\lambda_1+\lambda_{2}\lambda_1 R_1 & \lambda_{2}+\lambda_1-\lambda_{2}\lambda_1 R_1\end{pmatrix}}_{\text{Interface }i=1} \underbrace{\begin{pmatrix} 0\\1\end{pmatrix}}_{\text{BC}}
\end{equation}

As we add layers to the system a recursive product develops as follow:
\begin{equation}\label{eq_59}
\begin{pmatrix} \mathcal{B}^+\\\mathcal{B}^-\end{pmatrix}=\Lambda_1\,\left(\prod_{j=2}^{n-1}\,U_{j}\,\Lambda_{j}\right)\begin{pmatrix} 0\\1\end{pmatrix}
\end{equation}

\noindent
with the interface ---$\Lambda_{j}$--- and transfer ---$U_{j}$--- matrices defined as:
\begin{align}
&\Lambda_{j} = \frac{1}{2\lambda_{j+1}}\begin{pmatrix} \lambda_{j+1}+\lambda_j+\lambda_{j+1}\lambda_j R_j & \lambda_{j+1}-\lambda_j-\lambda_{j+1}\lambda_j R_j\\\lambda_{j+1}-\lambda_j+\lambda_{j+1}\lambda_j R_j & \lambda_{j+1}+\lambda_j-\lambda_{j+1}\lambda_j R_j\end{pmatrix}\label{eq_59b}\,,\\
&U_{j} = \begin{pmatrix}\exp{(-u_{j}d_{j})}&0\\0&\exp{(u_{j}d_{j})}\end{pmatrix}\label{eq_59c}
\end{align}

\noindent
where $u_j=(\lambda_j/k_{\text{y},j})$ is the transfer wave vector, $d_j$ is the thickness of layer $j$, and the dependence of all other parameters on $\eta$ is implied. Regardless of the number of layers, $\mathcal{A}^+(\eta)=1/2$, and $\mathcal{A}^-(\eta)=1/2$ (i.e., the heater is on the top of a the stack, otherwise see Bauer \textit{et al.} 2014). When $j=nw$, then $\mathcal{B}^+_1(\eta)=0$, and $\mathcal{B}^-_1(\eta)=1$ ---considering semi-infinite boundary condition---. The $R_j$ parameter it is defined as the thermal boundary resistance at an interface $j$, known as the Kapitza resistance~\cite{bauer2014}.

Comparing between the Cahill model, Meijer-G function, Borca-Tarsiuc model and the matrix formalism for a heater on a substrate system ---Fig.~\ref{fig_2}---, we can see that the numerical integration solutions to the equation agree each other. As pointed out by Cahill~\cite{cahill1990}, the integratin is an approximation, which actually we see that largely differs from the Meijer-G solution in the low frequency range, particularly the real part, with an absolute difference up to 0.5 in the normalized scale. The difference for the out of phase part, is narrowed for all solutions. We must use the analytical model whenever experimenting with heater on a substrate, leaving to any of the recursive formalism for multilayer systems. We need to consider these results since the thermal conductivity is obtained from the real part of the temperature rise. For the silicon example, the error becomes negligible as the frequency is over \SI{10}{\hertz}.

\begin{figure}[t]
    \begin{center}
        \includegraphics[scale=0.47]{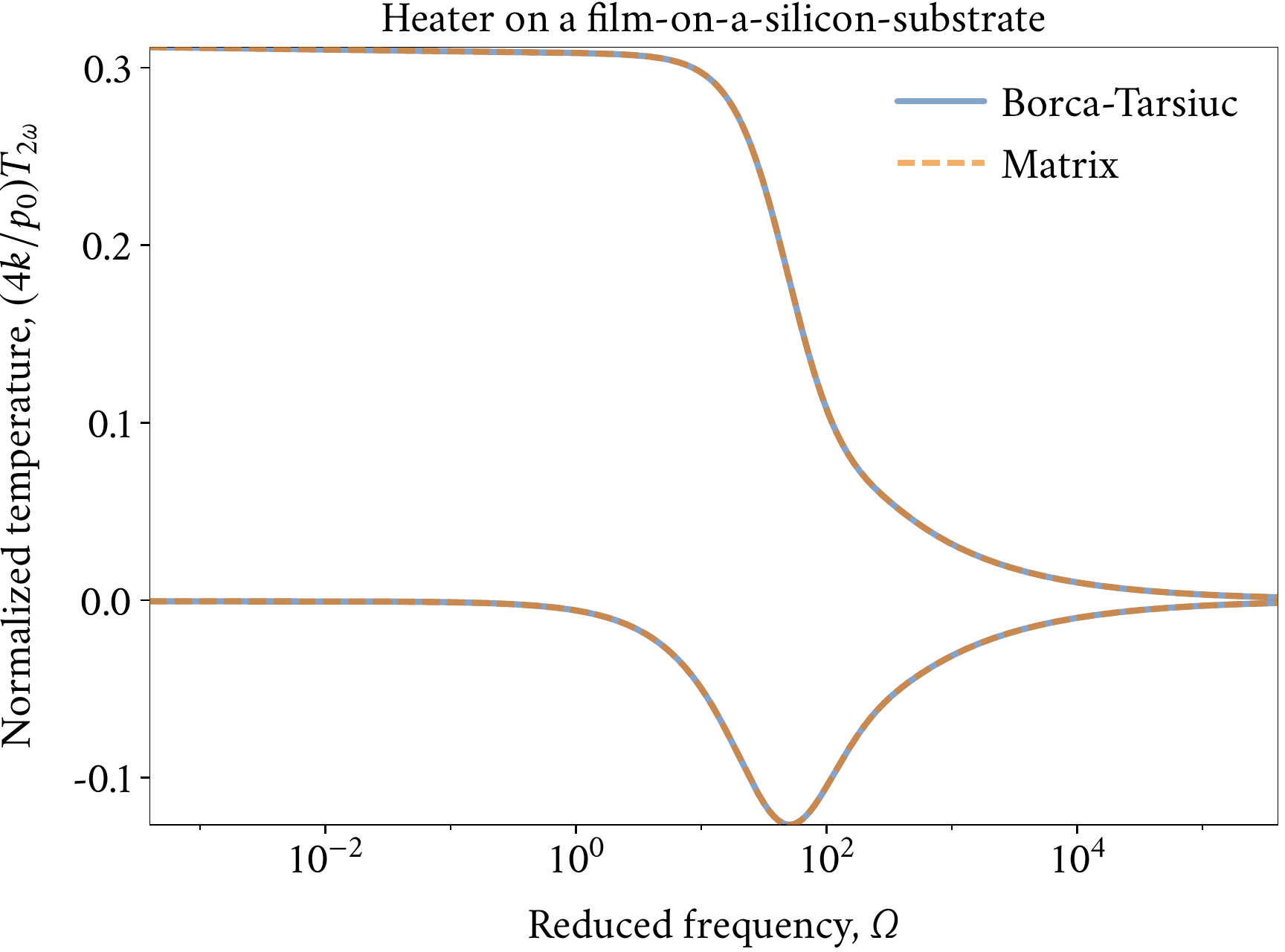}
        \vspace{0.3cm}
        \caption{$T_{2\omega}$ spectrum of an isotropic \SI{1}{\micro\meter}-thick film on bulk silicon, with $k_y=\SI{0.1}{\watt\per\meter\per\kelvin}$ keeping the rest of the values as calculated in Fig.~\ref{fig_2}, calculated with Eqs.~\eqref{eq_55} ---Borca-Tarsiuc--- and \eqref{eq_57} ---Matrix formalism---.\hfill{ }}
        \label{fig_3}
        \vspace{-0.5cm}
    \end{center}
\end{figure}

For a two layer system, the real part of the temperature rise of Borca-Tarsiuc model matches that of the matrix formalism ---Fig.~\ref{fig_3}---. There is no close solution ---the author is aware of--- of the integrals involving recurrent multilayer information.

\section{Effect of the metal interface on the system}

In order to consider the heat conduction effects inside the heater, its heat capacitance $(\rho_{\text{h}}\,c_{\text{h}})$ and thickness $d_{\text{h}}$, and the thermal resistance $R_{\text{h}}$ between the heater and the film in contact with the heater should be included into the model. $\rho_{\text{h}}$ and $c_{\text{h}}$ are the density and the specific heat capacity of the heater, respectively. Under these conditions the complex temperature rise of the heater is given by the following expression:
\begin{equation}\label{eq_62}
T_{2\omega,\text{h}} = \frac{T_{2\omega} + R_{\text{h}}\, \left(\dfrac{p}{2 b}\right)}{1 + i\, 2\, \omega\, \rho_{\text{h}}c_{\text{h}}d_{\text{h}} \left(R_{\text{h}} + T_{2\omega}\, \dfrac{2 b}{p}\right)}\,.
\end{equation}

\section{Effects of black-body radiation and convection}

\begin{figure}[t]
    \begin{center}
        \includegraphics[scale=0.44]{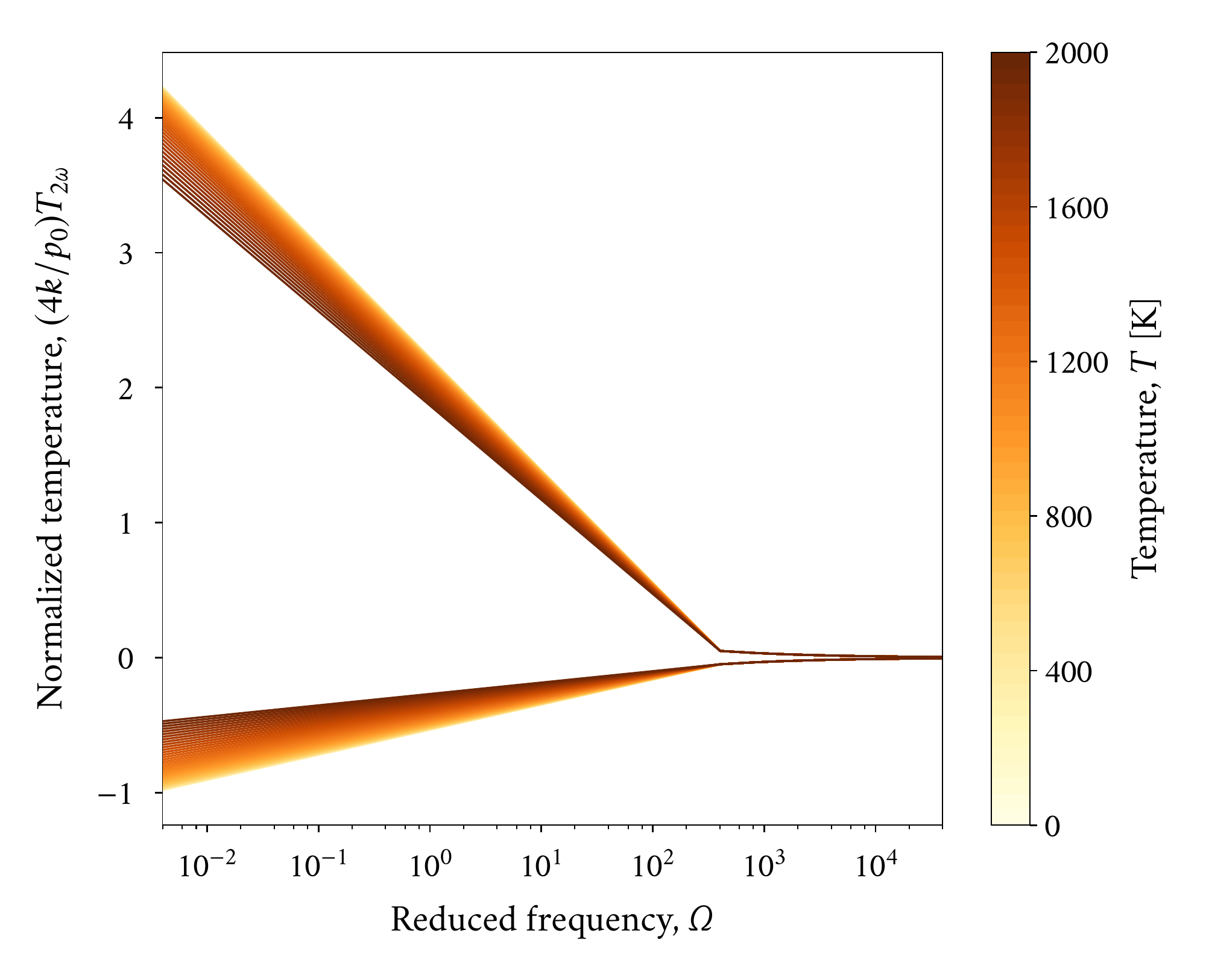}
        \caption{Temperature rise variation for different temperatures in the black-body radiation effect ---Eqs.~\eqref{eq_63}---, taking $\epsilon=1$.\hfill{ }}
        \label{fig_radiation}
        \vspace{-0.5cm}
    \end{center}
\end{figure}

The power is lost by radiation from the surface of the sample at position $x$ according to: $4\epsilon\sigma T^3 \mathit{\Delta} T(x)$. To evaluate the influence of this radiative term, to Eq.~\eqref{eq_48}, Cahill has proposed a corrected factor to the average temperature rise given by~\cite{cahill1990, cahill1987}:
\begin{equation}\label{eq_63}
T_{2\omega} = \frac{p_0}{\pi}\int_0^\infty \frac{\sin^2(\eta b)}{(\eta b)^2\left(k\sqrt{\eta^2+q^2}+2\epsilon\sigma T^3\right)}\text{d}\eta\,,
\end{equation}

\noindent
where $\epsilon$ is the emissivity, and $\sigma=\SI{5.67d-8}{\watt\per\square\meter\per\kelvin\tothe{4}}$ is the Stefan-Boltzmann constant. Fig.~\ref{fig_radiation} shows a numerical analysis of the effect of the temperature on a substrate of \SI{0.1}{\watt\per\meter\per\kelvin}, with thermal diffusivity \SI{6.8d-8}{\square\meter\per\second}, $b=\SI{6.25}{\micro\meter}$ and $l = \SI{1}{\milli\meter}$. The effects of radiation on the temperature oscillations are shown for several temperature values using an emissivity of 1. The influence of radiation is small, corresponding to an error in determining thermal conductivity less than 2\%~\cite{cahill1990, cahill1987}.

Dames explained that the combined effects can be disregarded given some considerations are taken into account for the experimental system~\cite{dames2013}. Values for the microscopic heater strips as used in 3$\omega$ experiments are not readily apparent in the literature but will be subject to two competing effects. The narrower heater width tends to increase the convection heat transfer coefficient $h_{\text{conv}}$, while the surrounding unheated substrate tends to impede air flow and reduce $h_{\text{conv}}$. The relative impact of heat losses is
\begin{equation}\label{eq_64}
\frac{Q_{\text{rad+conv}}}{Q_{\text{cond}}} \approx \frac{h\,q_{\text{s}}}{2\,k_{\text{s}}}\,,
\end{equation}

\noindent
where $Q_{\text{rad}}$, $Q_{\text{conv}}$ and $Q_{\text{cond}}$ are the radiation, convective and conduction heat power, respectively, while $q_{\text{s}}$ and $k_{\text{s}}$ are the substrate thermal wavelength and thermal conductivity, respectively, and $h=h_{\text{conv}}+h_{\text{rad}}$ is the total heat transfer coefficient ---convective plus radiative---. For a typical experiment with $q_{\text{s}}\approx\SI{100}{\micro\meter}$ and $k_{\text{s}}\approx \SI{100}{\watt\per\meter\per\kelvin}$, the losses are <1\% if $h<\SI{20,000}{\watt\per\meter\per\kelvin}$, which as noted above is very easily satisfied. In contrast, DC measurements on larger samples with characteristic lengths at the centimeter scale would reduce the threshold $h$ to the low hundreds of \si{\watt\per\meter\tothe{2}\per\kelvin}. \finofdoc

\linefindoc
\bibliographystyle{unsrt}

\end{document}